\documentclass[]{aa}
\usepackage{epsfig,amsfonts,amssymb,graphicx,fancyheadings,caption} 
\usepackage{multirow}
\usepackage{natbib}
\begin{document} 


\title{The OGLE-II event sc5\_2859 : a Classical Nova outburst?}
\author{  
C.~Afonso\inst{1,3},
J.F.~Glicenstein\inst{1},
A.~Gould\inst{2},
M.C~Smith\inst{4},
R.M.Wagner\inst{2},
J.N.~Albert\inst{5},
J.~Andersen\inst{8},
R.~Ansari\inst{5},
\'E.~Aubourg\inst{1},
P.~Bareyre\inst{1},
J.P.~Beaulieu\inst{6},
G.~Blanc\inst{1,12},
X.~Charlot\inst{1},
C.~Coutures\inst{1,6},
R.~Ferlet\inst{6},
P.Fouqu\'e\inst{9,10},
B.~Goldman\inst{1,3},
D.~Graff\inst{1},
M.~Gros\inst{1},
J.~Haissinski\inst{5},
C. Hamadache\inst{1,13},
J.~de Kat\inst{1},
L.LeGuillou\inst{1,11},
\'E.~Lesquoy\inst{1,6},
C.~Loup\inst{6},
C.~Magneville\inst{1},
J.B.~Marquette\inst{6},
\'E.~Maurice\inst{7},
A.~Maury\inst{10},
A.~Milsztajn \inst{1},
M.~Moniez\inst{5},
N.~Palanque-Delabrouille\inst{1},
O.~Perdereau\inst{5},
L.~Pr\'evot\inst{7},
Y.R. Rahal\inst{5},
J.~Rich\inst{1},
M.~Spiro\inst{1},
P.~Tisserand\inst{1},
A.~Vidal-Madjar\inst{6},
L.~Vigroux\inst{1,6},
S.~Zylberajch\inst{1}
}    
\institute{
CEA, DSM, DAPNIA,
Centre d'\'Etudes de Saclay, 91191 Gif-sur-Yvette Cedex, France
\and
Department of Astronomy, Ohio State University, Columbus,
OH 43210, U.S.A.
\and 
Max-Planck fuer Astronomie, Koenigsthul  17, 69117  Heidelberg, Germany
\and
Kapteyn Astronomical Institute,
University of Groningen,
P.O. Box 800, 9700 AV Groningen, the Netherlands
\and
Laboratoire de l'Acc\'{e}l\'{e}rateur Lin\'{e}aire,
IN2P3 CNRS, Universit\'e de Paris-Sud, 91405 Orsay Cedex, France
\and
Institut d'Astrophysique de Paris, INSU CNRS,
98~bis Boulevard Arago, 75014 Paris, France
\and
Observatoire de Marseille,
2 place Le Verrier, 13248 Marseille Cedex 04, France
\and
Astronomical Observatory, Copenhagen University, Juliane Maries Vej 30,
2100 Copenhagen, Denmark
\and
Observatoire Midi-Pyr\'en\'ees, 14 av. E. Belin, 31400 Toulouse
\and
European Southern Observatory (ESO), Casilla 19001, Santiago 19, Chile
\and
Instituut voor Sterrenkunde,
Celestijnenlaan 200B, 3001 Leuven, Belgium 
\and
Osservatorio Astronomico di Padova, INAF,
Vicolo dell'Osservatorio 5,
35122 Padova,
Italia
\and
Physique Corpusculaire et Cosmologie, Collège de France, 11 pl. M. Berthelot, 75231 Paris Cedex 5, France
}
\offprints{glicens@hep.saclay.cea.fr}         
\date{Received;accepted}
\authorrunning{C. Afonso et al.}
\titlerunning{The OGLE-II event sc5\_2859 : a Nova outburst}

\def\lsim{{\lesssim}}
\def\au{{\rm AU}}
\def\etal{{et al.}}
\def\eros{{\sc eros}}
\def\macho{{\sc macho}}
\def\lmc{{\sc lmc}}
\def\smc{{\sc smc}}
\def\ie{{\em i.e.}}
\def\tempest%
{\begin{array}{ccc}
1 & 1 & 1 \\
1 & 1 & 1 \\
4 & 3 & 8
\end{array}}
\def\gsim{{{}_>\atop{}^{{}^\sim}}}
\def\lsim{{{}_<\atop{}^{{}^\sim}}}
\def\kms{{\rm km}\,{\rm s}^{-1}}
\def\kpc{{\rm kpc}}
\def\e{{\rm E}}
\def\rel{{\rm rel}}
\def\btheta{{\vec\theta}}
\def\bmu{{\vec\mu}}
\def\bpi{{\vec\pi}}
\def\Teff{{T_{\rm eff}}} 
\def\msun{\rm M_\odot} 

\abstract{The OGLE-II event sc5\_2859 
was previously identified as the third longest microlensing event ever 
observed. Additional photometric 
observations from the EROS (Exp\'erience de Recherche d'Objets Sombres) 
survey and spectroscopic observations of the candidate star
are used to test the microlensing hypothesis.The combined OGLE and EROS data
provide a high quality coverage of the light curve. The colour 
of the sc5\_2859 event is seen to change with time. A 
spectrum taken in 2003 exhibits a strong $H_{\alpha}$ emission line.
The additionnal data  
show that the OGLE-II sc5\_2859 event is actually a classical nova outburst.  
}  {}

\maketitle          

\section{Introduction} 
Gravitational microlensing is now a well established tool for astrophysics.  
Of the more than 2000 microlensing events discovered to date,
the great majority are in the direction of
the Galactic bulge.  The microlensing optical depth and 
the event timescale distribution obtained from these surveys
provide unique tools for the study of Galactic structure and the
mass functions of Galactic populations. Moreover, microlensing 
databases have proven to be an extremely 
rich source for the study of various classes of variable stars.  In 
particular, they hold out the potential to obtain nova 
light curves  with higher cadence and better precision than
is usually the case.  

The confusion between falling nova light curves and microlensing events 
is a potential problem for microlensing studies.  The  
event sc5\_2859, discussed in this paper, stands both as a concrete 
example of this confusion and as a warning about its impact on calculations
of the optical depth.  Nov\ae\ have much larger outburst amplitudes 
than most microlensing events. 
But they are also easily distinguished from microlensing
events from their very different light curves on the rising side of the
peak.  Hence, if these analyses excluded events with data points only 
over the descending part of the light curve, this potential confusion would 
have no practical effect on optical depth measurements.
This is not the case, however, for two published optical-depth measurements
\citep{alcock97,alcock00}, which included, respectively, one and two 
decline-only events with
long Einstein timescales $t_{\rm E}$.  Specifically,
event 124-A ($t_{\rm E}=76\,$days), for the first paper and
events 95-BLG-d11 ($t_{\rm E}=62\,$days) and 96-BLG-d14 ($t_{\rm E}=47\,$days)
for the second.  Here $t_{\rm E}$ is given by
\begin{equation}
t_{\rm E} = {\theta_{\rm E}\over \mu_{\rm rel}},\qquad
\theta_{\rm E} = \sqrt{\kappa M\pi_{\rel}},
\label{eqn:tedef}
\end{equation}
where $\theta_{\rm E}$ is the angular Einstein radius,
$\pi_{\rm rel}$ and $\mu_{\rm rel}$ are the lens-source relative
parallax and proper motion, $M$ is the lens mass, and 
$\kappa\equiv 4G/c^2\rm AU\sim 8.1\,{\rm mas}$/$M_\odot$.
Although these events do not represent the bulk of the optical depth, 
they contribute significantly to it, since each event contributes
$\propto t_{\rm E}$.

Three very long events with timescales greater than one year have been 
discovered in the OGLE-II and 
MACHO data sets: OGLE-1999-BUL-32/MACHO-99-BLG-22 \citep{Mao2002,Bennett2002,Agol2002}, 
OGLE-1999-BUL-19 \citep{Smith2002} and BUL\_SC5 244353/sc5\_2859 
\citep{Smith2003}. 
The last of these, sc5\_2859, was initially found by the OGLE-II collaboration 
and later identified by \citet{Smith2003} 
as a long duration parallax event with 
one of the longest
time scales ever observed,
$t_\e = 547.6^{+22.6}_{-2.8}$ days, and as one of the first examples of 
disc-disc microlensing. 

In this paper we present the light curve of the event 
sc5\_2859 including additional points
from the EROS~2 
and OGLE projects databases. 
The new light curve and complementary 
spectroscopic observations of the candidate reveal that the event is most 
probably a nova outburst.     

\section{Observational Data}\label{sec:observation}
\begin{figure}
\begin{center}   
\mbox{\epsfig{file=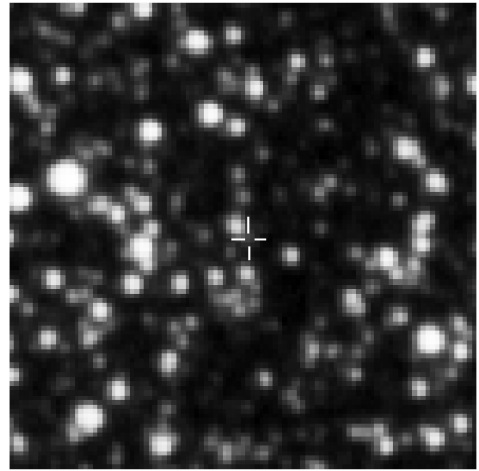,width=9.5 cm}}
\caption[]
{ EROS $I_{E}$ image showing the location of the source of sc5\_2859.
The image size is 1' x 1'. The location of the source is indicated by 
the cross. North is up and east is to the left.} 
\label{fig:findingchart}
\end{center}
\end{figure}

The event sc5\_2859 toward the Galactic bulge is a microlensing	candidate
from  the OGLE-II catalog. The light curve in the $I$ band is publicly 
available \footnote{ The Web address for the OGLE catalog is 
http://www.astro.princeton.edu/$\sim$wozniak/dia/lens/ }. The OGLE team did not 
observe the rising part of the light curve, whose
peak occurred just before the 1997 bulge observation campaign. A detailed 
description of the telescope and instrument can be found in 
\citet{Udalski1997}. The position of the source is
RA=17:50:36.09 and DEC=$-30$:01:46.6 ($l=359.\hskip-2pt^\circ 6235\hskip-2pt,
b=-1.\hskip-2pt^\circ 4930\hskip-2pt$). 
The location of the source is shown on Figure \ref{fig:findingchart}.

After \citet{Smith2003} presented a
microlensing model of the event, which included the effects of
microlensing parallax, the EROS~2 project searched its archival data 
for additional points that would allow a better constraint on the fit. 

\begin{figure}
\begin{center}   
\mbox{\epsfig{file=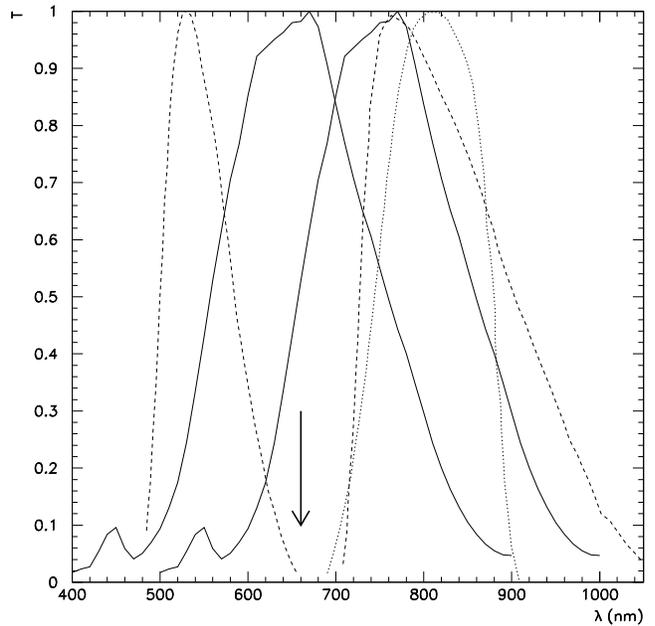,width=9.5 cm}}
\caption[]
{ EROS $V_{E}$ and $I_{E}$ passbands (solid lines). The broken lines show
OGLE V and I passbands and the dotted line is Landolt I passband. The arrow
shows the position of the H${\alpha}$ line.} 
\label{fig:passbands}
\end{center}
\end{figure}

The observations of the 
EROS~2 data were carried out at La Silla, Chile with the 1m MARLY telescope. 
The imaging was done simultaneously by two cameras (using a dichroic 
beam-splitter) composed of a mosaic of eight 
2K$\times$2K LORAL CCDs, with a field of view of 
$0.\hskip-2pt^\circ 7(\alpha)\times 1.\hskip-2pt^\circ 4(\delta)$ 
at an image scale of $0.\hskip-2pt''6$ per pixel.
The EROS filters are non-standard: one camera observes in the  
$V_{E}$ (420-720 nm) filter and the other in the 
$I_{E}$ filter (620-920 nm).
The EROS and OGLE passbands are shown in Figure \ref{fig:passbands}. Note that 
both EROS $I_{E}$ and $V_{E}$ are sensitive to possible contributions from the 
H${\alpha}$ line.
Approximate colour transformations  from EROS filters to standard 
colours are given by:
\begin{eqnarray}
V_E = 0.69\,V + 0.31\,I ,\quad
I_E = -0.01\,V + 1.01\,I,
\label{eq:transfcolor}
\end{eqnarray}
\citep{Regnault2000}.
However more precise colours were needed to compare the OGLE and EROS 
data sets. The colour transformations were found by comparing the approximate 
V,I values of equations (\ref{eq:transfcolor}) to the bulge V,I maps of OGLE
\footnote{ These maps are available at web address 
http://bulge.princeton.edu/$\sim$ogle/ogle2/bulge\_maps.html }.
More information about the EROS instrument can be obtained from \citet{BAU97}. 
The event sc5\_2859 is located in the EROS field Cg001 
close to the Galactic centre (see \citealt{Afonso2003} 
for a Galactic plane map of the EROS~2 bulge fields).  
   
\section{Light curve of sc5\_2859}

\begin{figure}[]
\begin{center}   
\mbox{\epsfig{file=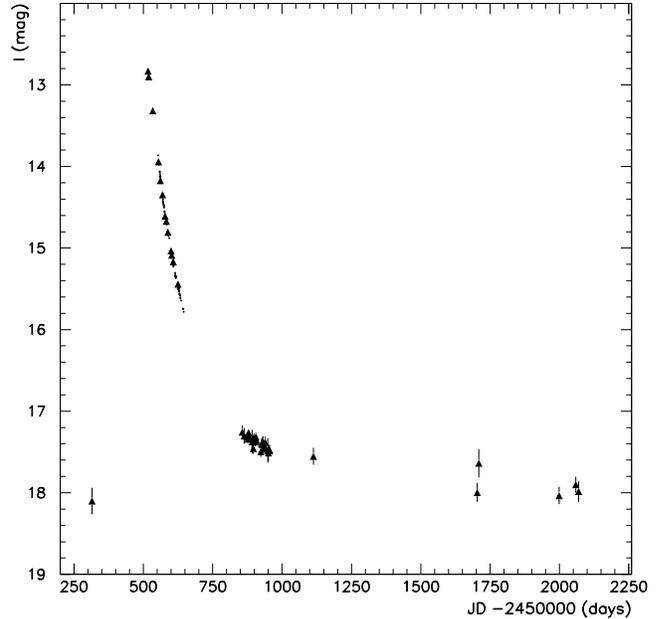,width=9.5 cm}}
\caption[]
{ Combined OGLE $I$ (dots) and EROS $I_{E}$  (triangles) light curve
for the OGLE sc5\_2859 event. Note the data point close to the baseline flux 
value at JD = 24500314.7. } 
\label{fig:comperosogle}
\end{center}
\end{figure}

The EROS data were processed by two different photometry pipelines. 
The PEIDA \citep{ANS96} PSF fitting photometry was run as part of the 
systematic photometry of EROS Galactic centre fields. The images were also 
processed with the ISIS \citep{ISIS} differential photometry package.
Differential photometry is  more accurate and yields 
errors that are at the level of photon 
noise. In this paper, we mostly use the better differential photometry.  
%
%
Figure \ref{fig:comperosogle} shows the OGLE $I$ and EROS $I_{E}$ light curves.
The two data sets have been  photometrically aligned by a linear 
transformation. The two data sets are in good overall agreement.

 The most striking feature of the $I$ light curve
(Fig.\ \ref{fig:comperosogle}) is the EROS $I_{E}$ data point 
at $JD = 24500314.71$. 
The $I_{E}$ flux at this date is at essentially the same level as it is 
1700 days later,
around $JD = 2452100$. The $I_{E}$ light curve rises from baseline to
peak in less than 200 days, but requires more than 400 days to 
return to baseline.  The strong asymmetry of the light curve 
is not easily explained by microlensing parallax.  
Unfortunately, no $V_{E}$ pre-peak data are available.  Although
some images were taken, their quality is very poor due to bad 
weather conditions.  Hence, the inference of strong asymmetry relies
on a single $I_E$ data point.  Nevertheless, visual inspection of the
image from which this datum was extracted reveals nothing anomalous.


\section{Colour evolution sc5\_2859} 

Another critical piece of information provided by the EROS data is the    
evolution of the colour of sc5\_2859 with time. EROS has obtained 
41 colour measurements over more than 4 years. In addition,
11 more colour measurements were made available to us by the 
OGLE collaboration \citep{Udalski2005}. The OGLE and EROS data are 
in apparent disagreement
regarding the colour (see Figures~\ref{fig:diaghr} and \ref{fig:colorvstime}). 
The EROS passbands
are wide and overlapping. The OGLE $V$ passband is close to Johnson $V,$ while 
the $I$ passband is non-standard. We will show in section \ref{sec:spectrum}
that the  spectrum of the sc5\_2859 object contains the H${\alpha}$ line
in emission. This does not affect either OGLE passband, but affects both the 
EROS passbands. The usual colour transformation do not apply in this case.  
However, we will now show that both OGLE and EROS data show evidence for
a chromaticity effect.  This is sufficient to rule out the microlensing
interpretation.

We now examine the colour changes from several different perspectives. 

\begin{figure}[]
\begin{center}   
\mbox{\epsfig{file=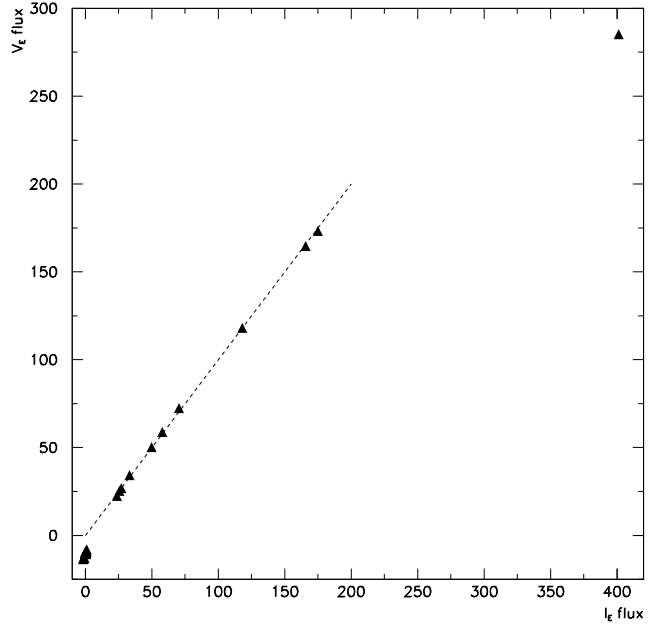,width=9.5cm}}
\caption[]
{ 
Differential flux in the $V_{E}$ band as a function of the  differential 
flux in the $I_{E}$ band. If sc5\_2859 were  a microlensing event, this curve would 
be a straight line, independently of the blending 
of the source. However, the curve deviates from a straight line (dotted line)
at both small and large fluxes.
} 
\label{fig:bluvsred}
\end{center}
\end{figure}

\begin{figure}
\begin{center}   
\mbox{\epsfig{file=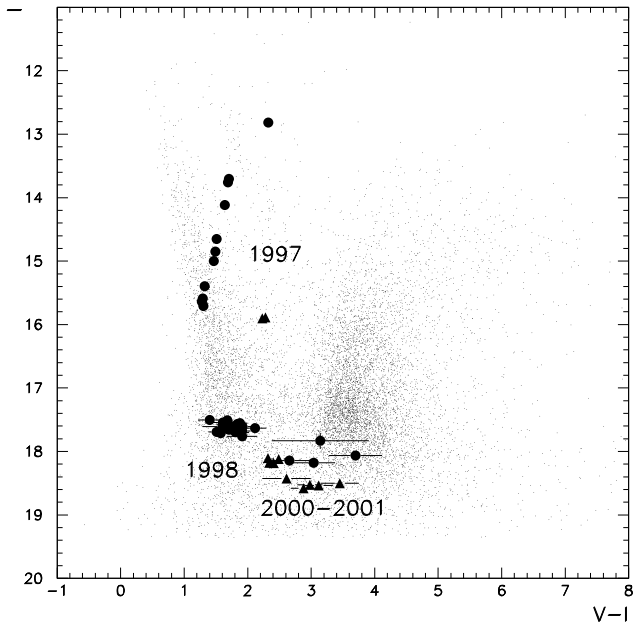,width=9.5cm}}
\caption[]
{ 
Evolution of the sc5\_2859 object in the local colour-magnitude diagram.
The dots are EROS2 data points and the triangle are OGLE data points.
The 1997 data can be interpreted as the end of the ``constant
bolometric luminosity'' phase (``phase 2'') and the 1998-2000-2001 data
as the ``white dwarf cooling phase''. 
} 
\label{fig:diaghr}
\end{center}
\end{figure}

\begin{figure}
\begin{center}   
\mbox{\epsfig{file=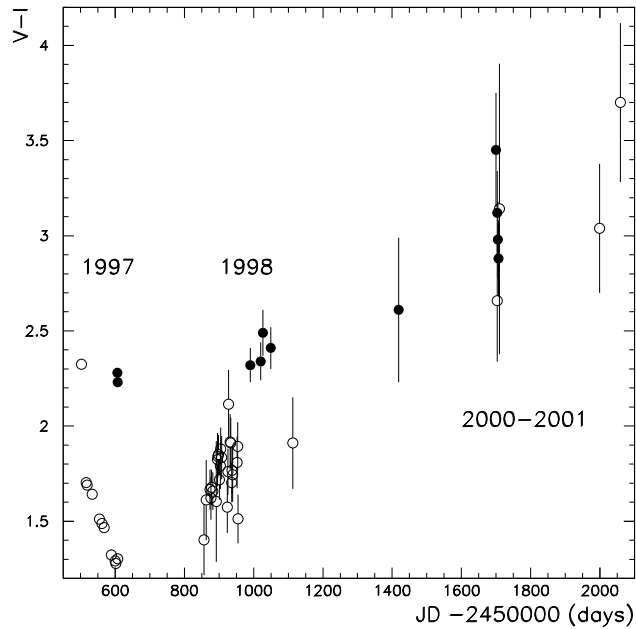,width=9.5cm}}
\caption[]
{ 
Evolution of the $V-I$ colour of the sc5\_2859 object with time.
The open circle are EROS data points and the filled circle are 
OGLE data points. 
Evolution of the EROS colour: 
1997: fades from $I=12.5$ to $I=15.7$.  1998: reddens from
$V-I=1.4$ to $V-I=2.0$.  2000-2001: near baseline at $V-I\sim 3$.
The colour measured by OGLE increases, but more slowly between 1997 and 2000.
} 
\label{fig:colorvstime}
\end{center}
\end{figure}

First, Figure \ref{fig:bluvsred} shows the differential $V_E$ flux
versus differential $I_E$ flux.  The track of points shows strong curvature.
By contrast, microlensing is achromatic and so (with a few exceptions 
to be noted below) predicts a strictly linear relation between
fluxes in different bandpasses, 
independently of the blending of the source (see e.g. Figure 4
of \citealt{Afonso2001}).  
The exceptions arise from cases in
which the source colour varies spatially, so that during the
course of the event the relative magnification of the blue and
red source material varies.  This can occur because of differential
limb darkening when the lens transits the source 
\citep{Witt1995,Yoo2004} or because both components of a binary
source are significantly magnified during the event, but by relative
factors that vary with time.  The first explanation is ruled out
by the lack of finite-source effects in the light curve, which would
always accompany significant colour changes.  The second explanation
is ruled out by the following argument.  If the event is modeled as
microlensing, then the peak observed flux corresponds to a magnification
$A>250$ (and could be even higher if the source were heavily blended),
corresponding to a projected lens-source separation 
(in units of $\theta_{\rm E}$) of $u\approx A^{-1}<0.004$, 
and hence a projected physical separation
$b<0.03\,{\rm AU}(\theta_{\rm E}/{\rm mas})$.  The putative binary
source would have to have a projected separation $r_\perp \la b$
if both sources were to contribute significantly to the light and
yet leave the light curve without pronounced bumps.  However, in
this case, the orbital period would be 
$P\la 1\,{\rm day} (\theta_{\rm E}/{\rm mas})^{3/2}$ which would
induce oscillatory (rather than the observed secular) colour changes
on the light curve, unless $\theta_{\rm E}$ were extremely large,
$\theta_{\rm E}\ga 50\,\rm mas$.  Such a large $\theta_{\rm E}$ 
would be implausible in its own right, but in addition 
(since $t_{\rm E}\sim 1\ \rm  year$),
this would imply $\mu_{\rm rel}\ga 50\,{\rm mas/yr}$, an impossibly 
large value.  In brief, Figure \ref{fig:bluvsred} essentially rules
out the microlensing interpretation of this event.

Second, in Figure \ref{fig:diaghr}, 
we show the evolution of the source colour and magnitude during the
event superposed on a $[(V-I),I]$ colour-magnitude diagram (CMD) composed of 
field sources within a $5'\times 5'$ square centered on the event.
EROS fluxes can be converted into standard $V,I$ magnitudes only 
at dates on which data in both passbands are available.
The differential fluxes are first 
converted into absolute fluxes by adding a constant. This constant is found  
by fitting a linear transformation (with slope $\sim 1$) between the ISIS and
the PEIDA fluxes. Finally, the fluxes are converted to standard magnitudes 
using the method described in section \ref{sec:observation}.
For most stars, this method gives a value close to the true standard 
$V,I$ magnitudes. For the source of sc5\_2859, the $V,I$ magnitudes obtained 
are not the standard magnitude, as explained earlier in this section.
The track of sc\_2859 in the CMD is close to the true track for OGLE points,
and it is basically the track that would be obtained in the 
EROS $[(V_{E}-I_{E}),I_{E}]$ CMD, since the transformation from
$(V_{E},I_{E})$ to $(V,I)$ is linear.

As mentioned above, microlensing events are in general achromatic, except
for blending scenarios in which the source star is blended by one or more stars
or in the presence of differential limb darkening. In the specific case
of event sc5\_2859, these exceptions are ruled out by the arguments presented
at the beginning of this section. Thus the colour change in the CMD
during the amplification of the light of the source star should depend
monotonically on the magnitude. This is clearly contradicted by Figure 
\ref{fig:diaghr}, where
one can see (for EROS points) a blueward decline 
(points in 1997) from (2.3,12.6) to (1.4,15.5)
and a clump at (1.5,17.7) representing a slower redward decline (points
in 1998), while the cluster of points just below the red clump are the
baseline (points in 2000-2001).
The OGLE points follow a qualitatively similar track in the CMD diagram.



The red giant clump is located at 
\begin{equation}
({(V-I)}_{clump},I_{clump}) \simeq (3.8, 17.4) 
\end{equation}
in the local colour-magnitude diagram. \citet{Yoo2004} argue that
the dereddened position of the red giant clump of the Galactic centre is 
located at 
\begin{equation}
[(V-I),I]^{GC}_0 \simeq (1.00, 14.32). 
\end{equation}
The reddening and local extinction of the clump is thus
\begin{equation}
[E(V-I),A_{I}]_{clump} \simeq (2.8,3.1),
\end{equation}
which implies a ratio total-to-selective extinction
$R_{VI} = A_V/E(V-I)=2.1$ which is consistent with the 
values found by \citet{Sumi2004} for bulge fields.

Third, we show in Figure \ref{fig:colorvstime} the time evolution
of the $V-I$ colour of the sc5\_2859 object.
The OGLE and EROS colours are obviously different.
Figure \ref{fig:colorvstime} shows that the colour 
changes with time, first decreasing (for EROS data points) by $\sim 0.7$ mags 
between $JD=2450500$ and $JD=2450700$, then steadily increasing by 
2 mags over the next 1400 days. The OGLE data points show a steady increase 
of the colour by $\simeq 0.7$ mags over 1200 days.

Apart from microlensing, the sc5\_2859 event has a natural explanation as a
cataclysmic variable (CV) outburst. 
This interpretation, which was also 
suggested by \citet{Cieslinski2003}, can be tested directly on the data and 
also by taking spectra of the candidate.

\section{Spectroscopic observations of the candidate}

\subsection{Observations}\label{sec:spectrum}
Optical spectroscopy of the nova candidate was obtained with the Red Channel 
CCD spectrograph on the 6.5m MMT of the Smithsonian Institution and the 
University of Arizona on 2003 July 8.15 UT.  Acquisition of the object on 
the 1\arcsec\ wide entrance slit was accomplished by performing a blind 
offset from an anonymous nearby field star located at 
$\alpha$ = 17:50:37.95 and $\delta$ = $-30$:01:39.4 (J2000), which was 
easily visible using the TV acquisition system (limiting magnitude $\sim$21.5).
  A 150 grooves mm$^{-1}$ grating was employed that covered the spectral 
region 4000-9200 \AA\ at a resolution of 19 \AA. However, the signal-to-noise 
ratio (S/N) is poor below 5500 \AA\ and longward of 7500 \AA.  Two 1200 s 
exposures were obtained at starting airmasses of 2.16 and 2.27 respectively.  
The seeing averaged 1$\farcs$6 at these airmasses as measured by the FWHM of 
anonymous field stars on the slit.  The transparency was of photometric 
quality based on spectrograph throughput measurements obtained earlier in the 
night.  The spectrum of a HeArNe lamp was obtained immediately after the 
target spectra for accurate wavelength calibration.  Three spectra were 
obtained near the end of the night of the standard star BD+28$\degr$ 4211 
through a 5$\arcsec$ wide entrance slit in order to correct the target 
spectra for instrumental response.  Bias and flatfield spectra were obtained 
to facilitate the data reduction.  The spectra were reduced, extracted, and 
calibrated using the IRAF software package.

\subsection{Results}

\begin{figure}
\mbox{\epsfig{file=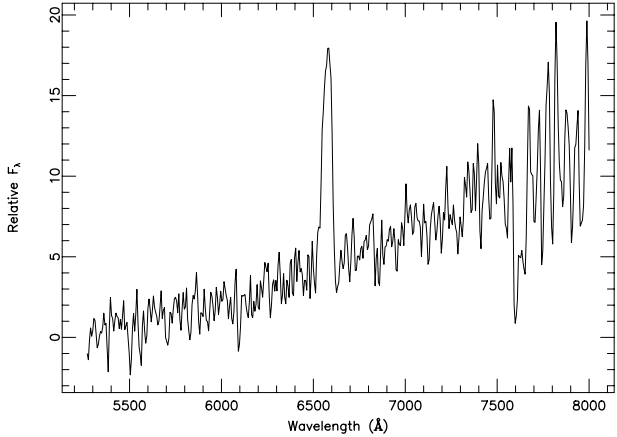,width=9.5cm}}
\caption{Spectrum of the progenitor of the OGLE-II event SC5\_2859 obtained 
with the 6.5m MMT on 2003 July 8.  The spectrum is dominated by a strong and 
broad H$\alpha$ emission line superposed on a reddened continuum.  The 
spectrum resembles that of some cataclysmic variables.}
\label{fig:spectrum}
\end{figure}

The extracted and calibrated spectrum of the nova candidate covering the 
5500-7500 \AA\ region is shown in Figure \ref{fig:spectrum}.  The spectrum 
exhibits a strong and broad H$\alpha$ emission line superposed on a reddened 
and noisy featureless continuum.  The equivalent width of H$\alpha$ emission 
is 176 \AA\ while its FWHM is $\simeq$2400 km s$^{-1}$.  The S/N is 
too poor in the blue to confirm the presence of any other other emission 
lines such as H$\beta$, \ion{He}{II} 4686 \AA, \ion{He}{I} 5876 \AA, and the 
\ion{C}{III}-\ion{N}{III} blend at 4650 \AA.  Similarly, strong interstellar 
absorption bands and absorption lines or bands of \ion{Na}{I} D and TiO 
characteristic of a late-type companion star do not appear to be present,
which is not surprising given our poor S/N.  The data do not permit a 
significant limit to be placed on the reddening
along the line of sight based on the strength of 
interstellar absorption bands.

\section{Discussion}\label{sec:discussion}

The spectrum of the stellar object associated the OGLE-II event 
SC5\_2859 resembles a reddened cataclysmic variable (CV) star.  
These systems consist of a Roche-lobe filling main sequence star and a 
white dwarf accreting material through an accretion disc surrounding 
the compact object.  The optical light is generally dominated by strong 
emission lines of the Balmer series of H, \ion{He}{I}, \ion{He}{II}, and 
the \ion{C}{III}/\ion{N}{III} Bowen blend at 4640-4640 \AA\ from the bright 
accretion disc.  A comprehensive review of the properties of CVs is given 
by \cite{Warner1995}. In comparing this event 
with other CVs, the great strength of 
H$\alpha$ emission and its breadth are both consistent with a system viewed 
at relatively high orbital inclination, close to 90\degr.

The presence of a CV associated with the outburst as measured by the OGLE and
EROS experiments offers 
a natural explanation for the nature of the observed event.  
CVs are the progenitors of both dwarf novae and classical novae outbursts.  
The spectrum of this object is certainly consistent with other CVs that 
undergo dwarf nova outbursts such as U Gem and SS Cyg.  The quiescent 
spectrum of these objects is dominated by strong Balmer emission lines 
accompanied by the lack of strong \ion{He}{I} and \ion{He}{II} emission lines 
relative to H$\beta$.  In addition, the \ion{C}{III}/\ion{N}{III} blend is 
nearly always weak or absent in the quiescent spectra of dwarf novae. 
However, dwarf nova outbursts are relatively fast events with amplitudes of 
$\simeq$2-4 magnitudes and durations of less than $\sim$20 days.  
In addition, they show a recurrence time scale $\sim$100 days or less.  
The light curve of OGLE SC5\_2859 is inconsistent with the dwarf nova 
interpretation since the duration of the outburst is considerably longer 
and it is not observed to recur over the $\sim$1300 days of monitoring.

The most likely interpretation of the 
 sc5\_2859 event is that it is a 
classical nova outburst.  Classical novae occur in CV binary systems as a 
result of a thermonuclear runaway on the surface of the white dwarf primary 
star. 
The time evolution of classical nov\ae\ outbursts can be separated into 4 
phases. These phases are 1) the rise to maximum visual light, 2) 
a period of constant bolometric luminosity, 3) a period of cooling of
the white dwarf and 4) the re-establishment of accretion from the 
secondary. The evolutionary track of classical nov\ae\ remnants is shown in 
Figure 5.1 of \citet{Classical Novae}.
This figure uses nova DQ Her 1934, for which the transition between phases 
2 and 3 occurs at $M_{V} \simeq 7.$ The $V$ magnitude changes by more than 
10 units during phase 2. 
 The  track of the
sc5\_2859 object in the local $(V-I,I)$ CMD is displayed
in Figure \ref{fig:diaghr}.
 The 1997 data may  be identified with the end of the 
``constant bolometric luminosity''
period and the 1998-2000-2001 data with the "white dwarf cooling phase".
The white dwarf cooling phase thus occurs after the turning point in
the colour evolution around $JD=2450860$ days.

\begin{figure}
\begin{center}   
\mbox{\epsfig{file=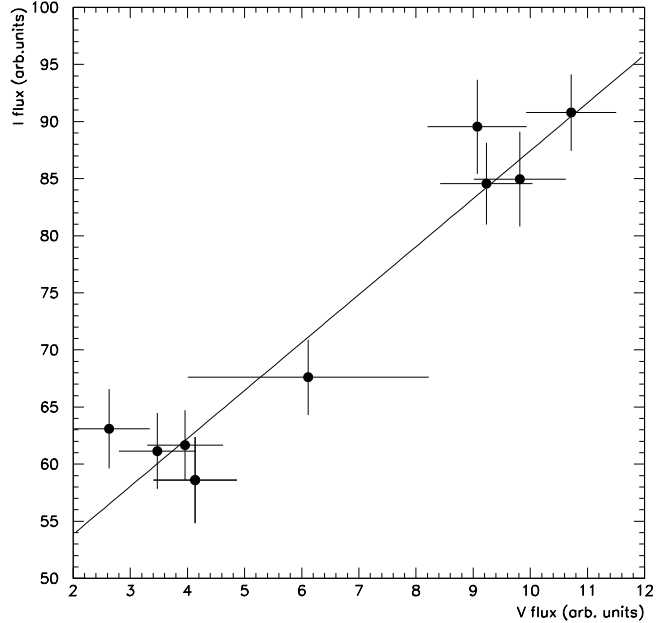,width=9.5cm}}
\caption[]
{ 
$I_{\rm OGLE}$ flux versus $V_{\rm OGLE}$ flux during the ``white dwarf cooling phase''.
The $I_{\rm OGLE}$ flux is a linear function of the $V_{\rm OGLE}$ flux because the 
$V-I$ colour of the dwarf is almost independent of temperature in the high 
temperature limit. 
} 
\label{fig:vvscoul}
\end{center}
\end{figure}

We now find the distance of the nova from the evolution of colour versus time.
For this, we use only OGLE colours, which are more reliable, since they
are not affected by the H${\alpha}$ line. 
Figure \ref{fig:vvscoul} shows a plot of $I_{\rm OGLE}$ flux versus 
$V_{\rm OGLE}$ flux during the ``white dwarf cooling phase''. 
This plot is very well fitted by a straight line 
with slope  $s=4.2 \pm 0.4.$
A possible interpretation is as follows: the flux comes from
the blend of two objects: one with $I$ and $V$
fluxes that are constant in time and another with $I$ and $V$ fluxes
decreasing with time, 
but with constant $V-I$ colour.
Translating the value of the slope $s$ in term of 
$V-I$ colour gives: ${(V-I)}^{\mbox{WD}} = 1.56.$
 
A high-temperature white dwarf 
cooling is expected to have an almost constant $V-I$ colour. 
This is because the spectrum 
of a high-temperature white dwarf cooling changes mainly in the UV. This 
qualitative argument is supported  by the calculations of 
\citet{Bergeron95}. They find that the $V-I$ colour of a hydrogen-atmosphere
white dwarf with $\log g = 8$ and temperature $> 30000$ K should be within 
0.05 mags of the asymptotic value ${(V-I)}^{H} = -0.36$  
(see also their Fig.~6). 
Using the theoretical value   ${(V-I)}^{H} = -0.36,$ one finds
for sc5\_2859, $E(V-I)= 1.56 - (-0.36)= 1.92.$ If a helium atmosphere
model is used instead, one gets $E(V-I)=1.82.$
These correspond respectively to 68.5\% and 65\% of the dust column
to the Galactic centre.


The projected height of the field is 4.0 deg $\times$ 8 kpc = 560 pc,
while the scale height of dust is 130 pc. Hence, for the nova to lie 
in front of
68.5\% the dust, a hydrogen atmosphere white dwarf would have to be at 
a height of $z=-130\ {\rm pc}\,\ln 0.315=150.2 {\rm pc}$ and so at a distance of
$D^{WD}= 8\ \mbox{kpc}(150.2\ \mbox{pc}/560\ \mbox{pc})= 2140\ \mbox{pc}.$
A similar calculation gives a distance of 1950 pc for a helium atmosphere 
dwarf. Taking the average value as the distance, we find
\begin{equation}
D^{WD}= 2045\ \pm 95\ \pm 500 \mbox{pc,}
\end{equation}
where the first error represents the uncertainty for 
the H/He ambiguity of the WD composition and the second
error represents our rough estimate in the uncertainty
in estimating the distance from the relative extinction.

{ 
Around $JD=2451704$, the apparent magnitude of the sc5\_2859 object is
$V = 21.5 \pm 0.05.$
Recalling that the exinction to the nova is
$A_V = R_{VI} E(V-I) = 4$, the absolute $V$ magnitude of the
source (assuming it is a hydrogen atmosphere WD) is then
\begin{equation} 
M_{V} \mbox{(JD=2451704)} \simeq 5.8.
\end{equation}
This is compatible with the minimum $M_{V}$ values given in Table 6 of 
\citet{Downes2000}. 
 However, the $M_{V}$ value of sc\_2859 might be somewhat lower
than 5.8, since it is very
 likely that the source is blended.
}
Note that the OGLE collaboration
found that the source of sc5\_2859 had a very high apparent proper motion
which they interpreted as the effect of a blend \citep{Sumi2004b}.

Since the earliest EROS $V_{E}$ point has $V \simeq 15$, it is likely
that most of the early decline part 
of the light curve (interpreted as ``phase 2'') was missed. The speed class 
of the nova can only be guessed 
at by means of qualitative arguments. The ``transition part'' of the V light  
curve is smooth and steadily declining. This favours a fast or very fast nova. 
Other features give only weak constraints. The slope of the $V$ light curve
around  $JD=2450600$ is 
\begin{equation}
\frac{dV}{dt} \simeq 0.02\ \mbox{mag/day},
\end{equation}
but this is only a lower limit to the early-decline slope. A similar
constraint comes from the date of the first data point of the $I_{E}$ 
light curve, which is observed before the outburst. These arguments
favour  a fast nova outburst.

{
Another useful check of the nova hypothesis is the calculation of the 
H${\alpha}$ line flux. The ratio of the R band continuum flux to the
I band continuum flux can be estimated from the spectrum. Transforming
to magnitudes using table 2.1 of \citet{BinneyMerrifield}, one finds
that
\begin{equation} 
R^{cont} - I = 1.53
\end{equation}
around $JD=2452829,$ so that $R \simeq 20.$
The $R$ absorption is $A_{R} \simeq 3.$ The $R^{cont}$ value  
corresponds to a continuum flux
in the R band of $2.3\ 10^{32}$ erg/s if the sc5\_2859 object is 
located at 2.05 kpc. The ratio of the H${\alpha}$ line flux to the 
continuum R band flux is roughly 0.5, so the H${\alpha}$ line
flux is roughly $10^{32}$ erg/s. H${\alpha}$ line fluxes of respectively 
1,\ 3 and 4~$10^{32}$ erg/s are expected for very fast, 
fast and moderately fast, 6.5 year old classical nova outbursts, 
according to \citet{Downes2001}.
The observed H${\alpha}$ line flux is thus in agreement with the hypothesis 
of a fast or very fast nova.
}

The paragraphs above summarize the arguments in favour of the classical 
nova hypothesis.
It is also possible in principle that the 
sc5\_2859 event is related to the subclass of recurrent novae 
such as T CrB, RS Oph, V3890 Sgr, and V745 Sco. 
The lack of obvious spectral features arising from a late-type giant star 
in the red weakens this association, however, in spite of the similarity of 
the light curve to some other recurrent novae.  
Nevertheless, a search for previous 
outbursts in archival plate material might prove worthwhile.

\bigskip

\begin{acknowledgements}
  We thank A. Udalski for interesting comments and for making unpublished 
  OGLE data on sc5\_2859i available to us.  
  We are grateful to D.~Lacroix and the technical staff at the Observatoire
  de Haute Provence and A.~Baranne for their help in refurbishing
  the MARLY telescope and remounting it in La Silla. We are also
  grateful to the technical staff of ESO, La Silla for the support
  given to the EROS project. We thank J-F.~Lecointe and A.~Gomes for the
  assistance with the online computing. Work by A.~Gould was supported
  by NSF grant AST~042758. M.C.Smith acknowledges financial support
from the Netherlands Organisation for
Scientific Research.
\end{acknowledgements}

\end{document}